\documentclass[aps,prd,preprint,groupedaddress,showpacs,nofootinbib]{revtex4}

\usepackage{amsmath}
\usepackage{amssymb}
\usepackage{mathtools}
\usepackage{graphicx}

\begin{document}

\title{Nonminimal coupling and the cosmological constant problem}

\author{Dra\v zen~Glavan}
\email[]{d.glavan@uu.nl}

\author{Tomislav~Prokopec}
\email[]{t.prokopec@uu.nl}

\affiliation{Institute for Theoretical Physics, Spinoza Institute and Center 
for Extreme Matter and Emergent Phenomena, 
 Utrecht University,
Postbus 80.195, 3508 TD Utrecht, The Netherlands}

\date{\today}

\begin{abstract}
 We consider a universe with a positive effective cosmological constant and a nonminimally coupled
scalar field. When the coupling constant is negative,
 the scalar field exhibits linear growth at asymptotically late times, resulting in a decaying effective cosmological constant.
The Hubble rate in the Jordan frame reaches a self-similar solution,
$H=1/(\epsilon t)$, where the principal slow roll parameter $\epsilon$ depends on $\xi$,
reaching maximally $\epsilon=2$ (radiation era scaling) in the limit when $\xi\rightarrow -\infty$.
Similar results are found in the Einstein frame (E), with $H_E=1/(\epsilon_E t)$, but now 
$\epsilon_E \rightarrow 4/3$ as  $\xi\rightarrow -\infty$. Therefore in the presence 
of a nonminimally coupled scalar de Sitter is not any more an attractor, but instead (when $\xi<-1/2$)
the Universe settles in a decelerating phase.  Next we show that, when the scalar field $\phi$ decays to matter with $\epsilon_m>4/3$ at a rate $\Gamma\gg H$, the scaling changes to that of matter, 
$\epsilon\rightarrow \epsilon_m$,
and the energy density in the effective cosmological becomes a fixed fraction of the matter energy density,
$M_{\rm P}^2\Lambda_{E\rm eff}/\rho_m={\rm constant}$,
exhibiting thus an attractor behavior. While this may solve the (old) cosmological constant problem, it does not explain
dark energy. Provided one accepts tuning at the $1\%$ level, the vacuum energy of neutrinos can explain the observed
dark energy.

\end{abstract}


\maketitle

\section{Introduction}
\label{Instroduction}

 Here we consider a simple tensor-scalar theory of gravity,
originally considered by Jordan, Brans and Dicke~\cite{Jordan:1955,Brans:1961sx}
and generalized by Bergmann~\cite{Bergmann:1968ve} to what we today refer as tensor-scalar (TeS) theory of gravity. 
The JBD -- and more generally TeS --  theories were used as a testing ground
for simplest extensions of general relativity (for a review see Ref.~\cite{Will:2005va}).
In cosmology the more general class of TeS theories (that includes a potential) has been used to formulate
the Higgs field driven inflationary models~\cite{Salopek:1988qh,Bezrukov:2007ep,Bezrukov:2013fka,Prokopec:2014iya},
and to build models that explain late time dark energy from inflationary 
fluctuations~\cite{Glavan:2014uga,Glavan:2015}.
Here we investigate how a nonminimally coupled scalar can help in resolving the cosmological constant problem.

 The cosmological constant problem regards the huge discrepancy between natural theoretical value 
and the observed value. While quantum field theory predicts a huge value 
(of the order of the Planck scale), the observed value is consistent with that of the dark energy, which is 
more than 120 orders of magnitude smaller.

 This problem has already been discussed in literature in the context
of a nonminimally coupled scalar field, and it was observed that 
the mechanism works, but the price to pay is that the effective gravitational coupling 
constant goes to zero, which is not acceptable~\cite{Dolgov:1982gh,Ford:1987de,Weinberg:1988cp}. 
In this we point out that this is indeed the case when one 
considers the problem in the Jordan frame. However, the analysis in the Einstein
frame warrants further investigation, and that is precisely what we do here. 

 In section~\ref{Jordan frame analysis} we define the model and perform the analysis 
in the Jordan frame. Section~\ref{Einstein frame analysis} is 
devoted to the corresponding analysis in the Einstein frame. In section~\ref{Adding matter} 
we extend the analysis of  section~\ref{Einstein frame analysis} by coupling the scalar field to matter 
and observe that in the tight coupling regime the effective cosmological constant in the Einstein 
frame scales away as matter. Finally, in section~\ref{Discussion and conclusions}
we discuss the results and address possible issues and shortcomings.

 \section{Jordan frame analysis}
 \label{Jordan frame analysis}


 The scalar-tensor model we consider here is defined by the following action in the Jordan frame,
\begin{equation}
  S = \int d^4 x\sqrt{-g}\left(\frac12 F(\phi)R - M_{\rm P}^2\Lambda - \frac12 g^{\mu\nu}\partial_\mu\phi\partial_\nu\phi-V(\phi)\right)
\,,
\label{action}
\end{equation}
where $g ={\rm det}[g_{\mu\nu}]$,  $g^{\mu\nu}$ is the inverse of the metric tensor $ g_{\mu\nu}$.
For simplicity we take,
\begin{equation}
   F(\phi) = M_{\rm P}^2 - \xi\phi^2\,,\qquad  {\rm and}\quad  V=0
\,,
\label{F and V}
\end{equation}
where $\xi$ is the nonminimal coupling. In our sign conventions conformal coupling corresponds to $\xi_c=1/6$,
$M_{\rm P}^2=1/(8\pi G_N)$ and we work with natural units in which $\hbar = 1 = c$. For the metric
we take a cosmological, spatially flat, background,
\begin{equation}
 g_{\mu\nu} = {\rm diag}[-1,a^2(t),a^2(t), a^2(t)]
\,.
\label{metric}
\end{equation}

 The action~(\ref{action}) implies the following equations of motion
(see {\it e.g.}~Refs.~\cite{Fakir:1990eg,Weenink:2010rr}),
\begin{eqnarray}
&&\hskip -1cm
\ddot \phi + 3H\dot\phi + 6\xi(2H^2+\dot H)\phi = 0
\label{basic EOM:1}\\
\dot H \;&=& -\frac{1}{M_{\rm P}^2-\xi(1-6\xi)\phi^2}\left[\frac12(1-2\xi)\dot \phi^2+4\xi H\phi\dot\phi+12\xi^2H^2\phi^2\right]
\label{basic EOM:2}\\
H^2 &=& \frac{1}{3(M_{\rm P}^2-\xi\phi^2)}\left[M_{\rm P}^2\Lambda+\frac12\dot \phi^2+6\xi H\phi\dot\phi\right]
\,,
\label{basic EOM:3}
\end{eqnarray}
where $H=\dot a/a$ is the Jordan frame Hubble rate.. 
Equation~(\ref{basic EOM:3}) is the constraint equation, and taking its time derivative
gives a combination of the first two equations~(\ref{basic EOM:1}--\ref{basic EOM:2}), representing a non-trivial
 validity check of the solutions of Eqs.~(\ref{basic EOM:1}--\ref{basic EOM:3}).
The cosmological constant $\Lambda$ appears only in the constraint equation~(\ref{basic EOM:3})
 and in that respect does not directly affect the dynamical equations~(\ref{basic EOM:1}--\ref{basic EOM:2}).
The information about $\Lambda$ is introduced by the initial conditions. 

We assume that the field is initially in a homogeneous state with a small expectation value, $\phi_0\sim \sqrt{\Lambda}$.
A detailed analysis is required to properly answer the question how much the subsequent analysis depends on the homogeneous
initial conditions. The following heuristic arguments suggest that our results do not depend on the details of
initial conditions.  Since the early period is de Sitter, during which fluctuations tend to be exponentially damped
with time,~\footnote{This of course does not hold for very long wavelength fluctuations, for which
the physical momentum, $k/a<\sqrt{-12\xi}H$. These fluctuations will also grow exponentially,
but slower than the homogeneous field mode, and hence our results can be understood as a lower bound
on the duration of the initial inflationary period.}
we expect that our results remain robust for a rather broad set of initial conditions.
In order to properly understand the dynamics and duration of the initial inflationary period,
it is important to include the backreaction of created scalar (and gravitational) particles,
which is work in progress. We expect that the backreaction from quantum fluctuations will shorten
the initial inflationary period~\cite{Glavan:2014uga,Tsamis:1996qq}. 
To make a more quantitative statement a (pertubative) quantum analysis is required.
In this work we examine the problem without including quantum backreaction effects. 

At early stages of the evolution the cosmological
constant contribution dominates the right hand side of Eq.~(\ref{basic EOM:3}), 
and $\epsilon=-\dot H/H^2\approx 0$, such that 
we are in an approximately de Sitter space
From Eq.~(\ref{basic EOM:1}) we see that, for a negative $\xi$, $\phi$ is tachyonic and
therefore it exhibits exponential growth,
\begin{equation}
 \phi(t) \approx \phi_0 \exp\left(-4\xi H t\right)
\,,\qquad a(t) \approx a_0 \exp(Ht)\,,\qquad H\approx \sqrt{\frac{\Lambda}{3}} ={\rm const.}
\label{early stages:solution}
\end{equation}
This stage ends at the time $t_{\rm end}\simeq N_{\rm end} /H$ when
the scalar field develops the energy density comparable to that of the cosmological constant
(see Eq.~(\ref{basic EOM:3})). This happens when 
 the number of e-folds  of inflation $N(t)=\ln(a(t)/a_0)$ is about,
\begin{equation}
  N_{\rm end} \simeq -\frac{1}{8\xi}\ln\left(\frac{3M_{\rm P}^2}{16\xi^2\phi_0^2}\right)
\,,
\label{number of efolds}
\end{equation}
such that one can get a large number of e-folds either by choosing $\xi$ or $\phi_0$ very small. For example, when $\xi=-1$
one gets $N_{\rm end}\sim 50$ when $\phi_0=(\sqrt{3}/4)\exp(-240)M_{\rm P}\simeq 5\times 10^{-60}~{\rm eV}$,
an extremely small field value, implying that this choice of parameters does not result in a good model for primordial inflation.
In this initial period, the principal slow roll parameter $\epsilon$
is exponentially close to {\it zero}, hence this period is, to an excellent approximation, a de Sitter epoch.

\begin{figure}[hht]
\begin{minipage}[t]{.4\textwidth}
        \begin{center}
\includegraphics[width=7.2cm,height=6.5cm]{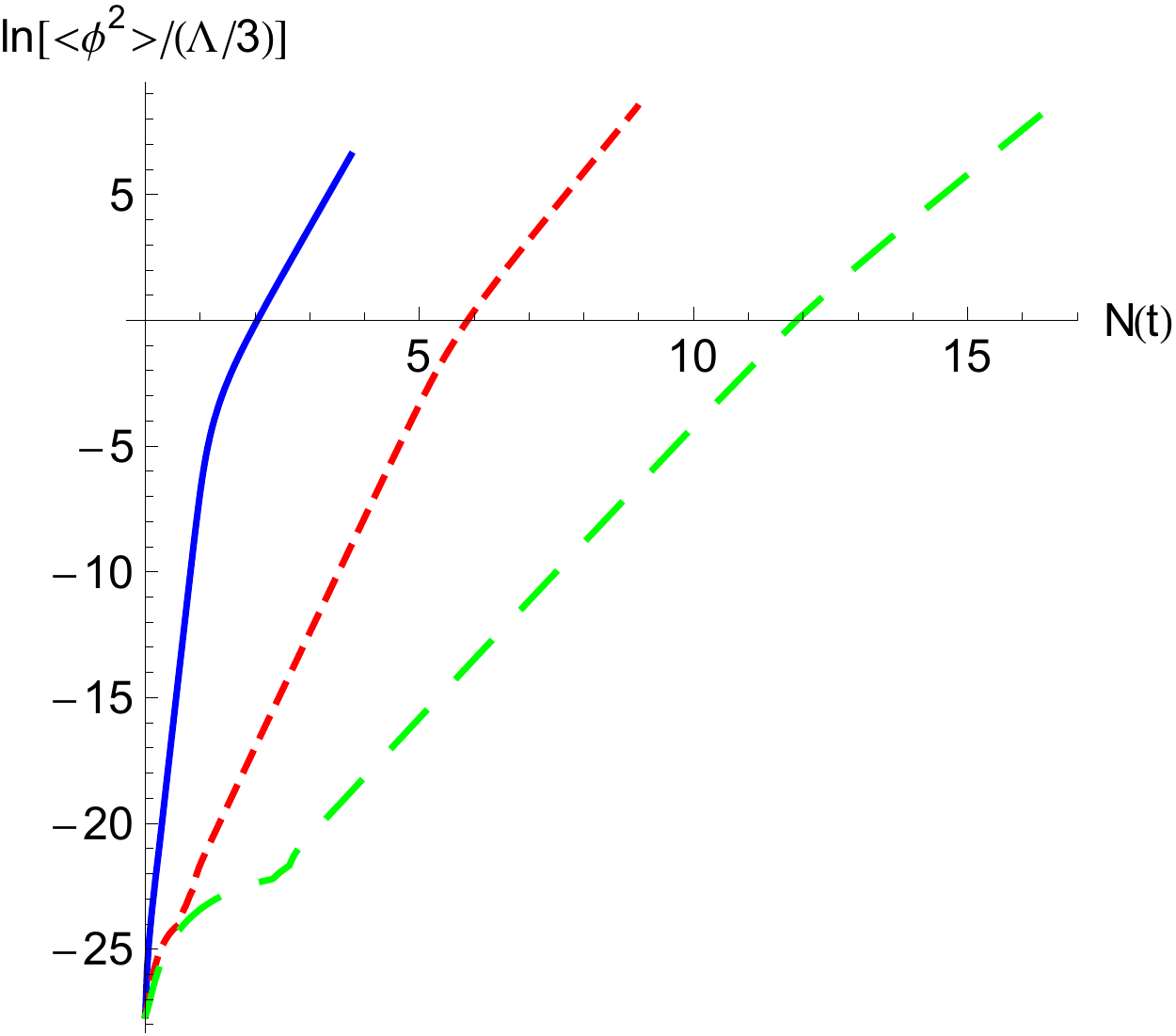}
\end{center}
    \end{minipage}
\hspace{1cm}
\begin{minipage}[t]{.4\textwidth}
        \begin{center}
\includegraphics[width=7.2cm,height=6.5cm]{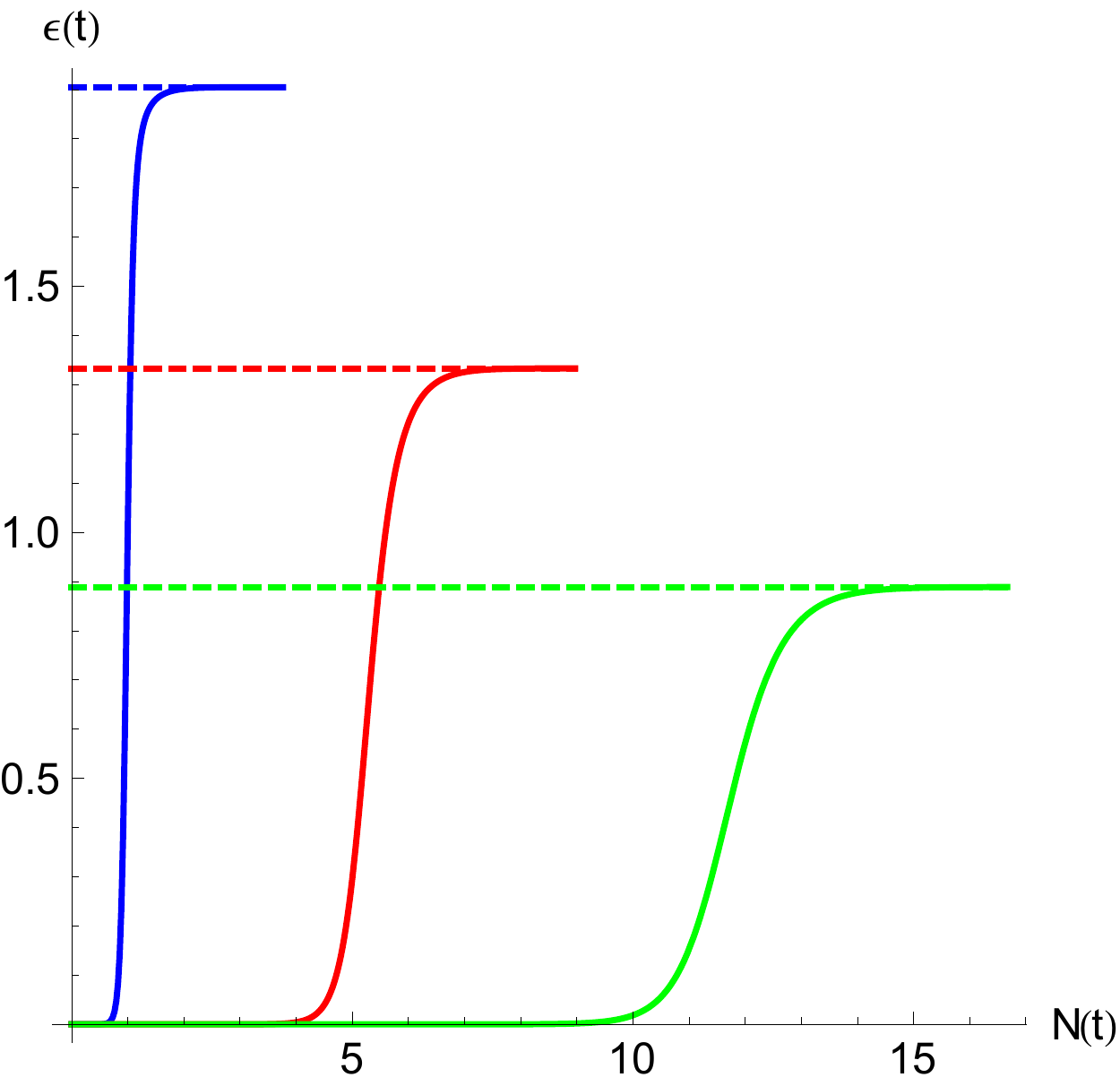}
\end{center}
    \end{minipage}
\caption{{\it Left panel:} $\ln(3\phi^2/\Lambda)$ as a function of the number of e-foldings.
Both in the quasi-de Sitter stage as well as in the late power-law expansion stage the field $\phi(t)$ grows
exponentially with the number of e-foldings, as explained in the main text. We show results for
$\xi = -10,-1$ and $\xi=-0.4$ (curves from left to right).
{\it Right panel:} $\epsilon=-\dot H/H^2$ as a function of the number of e-foldings. $\epsilon$ stays close to zero
during the early quasi-de Sitter, and transits to a constant value given in~(\ref{epsilon as function of xi})
 during the late-time attractor regime.
For $\xi = -10,-1$ and $\xi=-0.4$ (curves from left to right).}
\label{Phi2 and epsilon vs N}
\end{figure}

 After this initial quasi-de Sitter stage ends, a transitory stage sets it, after which the system enters an asymptotic regime.
We shall now show that there is an attractor solution in the asymptotic regime which is {\it not} de Sitter, {\it i.e.} during which
the (effective) cosmological constant relaxes rapidly to {\it zero}. To show that, let us make the following
scaling {\it Ansatz} which holds at late times,
\begin{equation}
 \phi(t)\rightarrow \dot\phi_0 t
\,,\qquad H(t) \rightarrow  \frac{1}{\epsilon t}
\,,
\label{late time solution}
\end{equation}
where $ \dot\phi_0$ and $\epsilon$ are constants. This {\it Ansatz} corresponds to a universe dominated by 
a perfect fluid with a constant equation of state parameter $w=p/\rho = -1+2\epsilon/3$, for which
the scale factor, $a(t)\rightarrow a_0(t/t_0)^{1/\epsilon}$.
Indeed, inserting~(\ref{late time solution}) into~(\ref{basic EOM:1}--\ref{basic EOM:2}) one obtains,
\begin{eqnarray}
&&  \epsilon = -\frac{4\xi}{1-2\xi}
\label{epsilon as function of xi}
\\
&&(1\!-\!2\xi)\epsilon^2+2\xi(5\!-\!6\xi)\epsilon+24\xi^2 = 0
\,.
\nonumber
\end{eqnarray}
From the two solutions to the latter equation, $\epsilon=-4\xi/(1-2\xi)$ and $\epsilon=-6\xi$,~\footnote{
That this is not a coincidence shows the analysis in $D$ spacetime dimensions, in which case
the two solutions are, $\xi=-D\xi/(1\!-\!2\xi)$ and $\xi=-2(D\!-\!1)\xi$, the former solution being 
the attractor in general $D$ spacetime dimensions.}
it is the former that
is consistent with the slow roll parameter~(\ref{epsilon as function of xi})
 implied by the scalar equation of motion. Numerical investigations,
an example of which is shown in figure~\ref{Phi2 and epsilon vs N}, shows
that, independently on the initial field value $\phi_0$,
the late time solution is given by~(\ref{late time solution}), implying that~(\ref{late time solution})
is an {\it attractor}. We conclude that, in a model with a scalar field with a negative nonminimal coupling,
a non-vanishing cosmological constant gets dynamically compensated by the field, and such a universe settles
in a power law expansion. From Eq.~(\ref{epsilon as function of xi}) we see that the late time solution
is accelerating ($\epsilon<1$) 
when $0>\xi>-1/2$ and decelerating ($\epsilon>1$) 
 when $\xi<-1/2$. In the limit when $\xi\rightarrow -\infty$,
$\epsilon\rightarrow 2$, which corresponds to a conformally coupled fluid (radiation). This conclusion
holds also in general $D$ space-time dimensions.
Inserting the {\it Ansatz}~(\ref{late time solution}) into the constraint equation~(\ref{basic EOM:3})
determines the late time rate of the scalar field growth,
\begin{equation}
\dot\phi_0^2= \frac{-8\xi}{(1\!-\!6\xi)(3\!-\!10\xi)}M_{\rm P}^2\Lambda
\label{dotphi0}
\end{equation}
In both periods scalar field grows exponentially with the number of e-foldings.
Indeed, at early times, $\phi\propto \exp[-4\xi N]$, while at late times,
$\phi\propto \exp(\epsilon N)= \exp[-4\xi N/(1\!-\!2\xi)]$, as can be seen from the left panel of 
figure~\ref{Phi2 and epsilon vs N}.

 The above analysis shows that a negatively coupled scalar relaxes a cosmological constant at a rate,
 \begin{equation}
  \Lambda_{\rm eff}(t) = \Lambda_0 a^{-2\epsilon}
\,,\qquad \epsilon = -\frac{4\xi}{1\!-\!2\xi}
  \,,
\label{Lambda eff}
\end{equation}
%
The question worth investigating is whether this can resolve the old cosmological constant problem~\cite{Weinberg:1988cp,Carroll:2000fy,Bousso:2007gp,Peebles:2002gy}:
{\it Why is the (observed) cosmological
constant so small when compared with its natural value suggested by quantum field
theory?} Here we assume that the cosmological constant generated by the vacuum fluctuations of quantum fields and by symmetry breaking
(such as the BEH mass generation mechanism) is of the order the electroweak scale,~\footnote{It
is often assumed that the natural value from quantum field fluctuations
is given by the Planck scale.
However, this cannot be so. Consider for simplicity the flat space case (since we are primarily discussing ultraviolet issues, 
due to the adiabaticity of the ultraviolet, the conclusions reached here are easily carried over to expanding backgrounds).
The Planckian value for the cosmological constant is obtained by setting the (physical) UV cutoff,
$\Lambda_{\rm UV}$ at the Planck scale, $\Lambda_{\rm UV}\sim m_{\rm P}$, $m_{\rm P} = \sqrt{8\pi} M_{\rm P}$.
In this case the one-loop contribution to the stress energy tensor can be described by an ideal fluid with the energy density
and pressure~\cite{Koksma:2011cq}, $\rho_{\rm UV} = \Lambda_{\rm UV}^4/(16\pi^2)=3p_{\rm UV}$,
implying an equation of state parameter of radiation,
$w_{\rm UV}=1/3$. Then the covariant energy conservation in a (homogenous) Universe dominated by such a vacuum energy,
$\rho_{\rm UV}+3H(\rho_{\rm UV}+p_{\rm UV})\simeq 0$
tells us that $\rho_{\rm UV}\propto 1/a^4$. That then implies that one either has to give up imposing
a physical momentum cutoff, or cutoff regularization altogether. Here we assume that cutoff regularization is incorrect,
since it violates the (observed) Lorentz symmetry of the (quantum) vacuum.
When a Lorentz symmetric regularisation is used~\cite{Koksma:2011cq}, one gets a universal (regularization independent)
result for the vacuum energy and pressure induced by (one-loop) vacuum fluctuations that is of the order the electroweak scale,
which we assume here to be the physical contribution to $\Lambda$ from (the vacuum fluctuations of) quantum fields.
Moreover, this result depends only logarithmically on the regularisation energy scale~\cite{Koksma:2011cq}, 
rendering this contribution stable under a change of the renormalization scale.}
\begin{equation}
  \rho_{\Lambda\rm EW}=M_{\rm P}^2\Lambda_{\rm EW}\sim [2\times 10^{2}~{\rm GeV}]^4
\;\Longrightarrow\;\Lambda_{\rm EW}\sim 4\times 10^{-28}~{\rm GeV}^2
\,.
\label{natural value cosmological constant}
\end{equation}
On the other hand, the cosmological constant corresponding to the observed dark energy is about
$\Lambda_{\rm DE} \simeq 4\times 10^{-84}~{\rm GeV}^2$, which is about $10^{56}$ times smaller.
This discrepancy between the natural (expected) value of the cosmological constant and the observed
one constitutes the cosmological constant problem. Based on the above analysis, one would be tempted to conclude
that the answer is positive. This is, however, not so for the following reason. Even though vacuum fluctuations of matter fields
that couple minimally (canonically) to gravity in the Jordan frame
will provide a large and approximately constant contribution to the cosmological constant, they will also feel a time dependent
effective Newton constant, $G_{\rm eff}(t) = G_N/[1-8\pi G_N\xi\phi^2(t)]$~\cite{Weinberg:1988cp},
and no such time dependence of the gravitational coupling strength
has been observed. This observation implies that the Jordan frame analysis cannot solve the (old) cosmological constant problem.
Let us, therefore, repeat the analysis in the Einstein frame.

 \section{Einstein frame analysis}
 \label{Einstein frame analysis}

\begin{figure}[hhh]
\includegraphics[width=10cm,height=8cm]{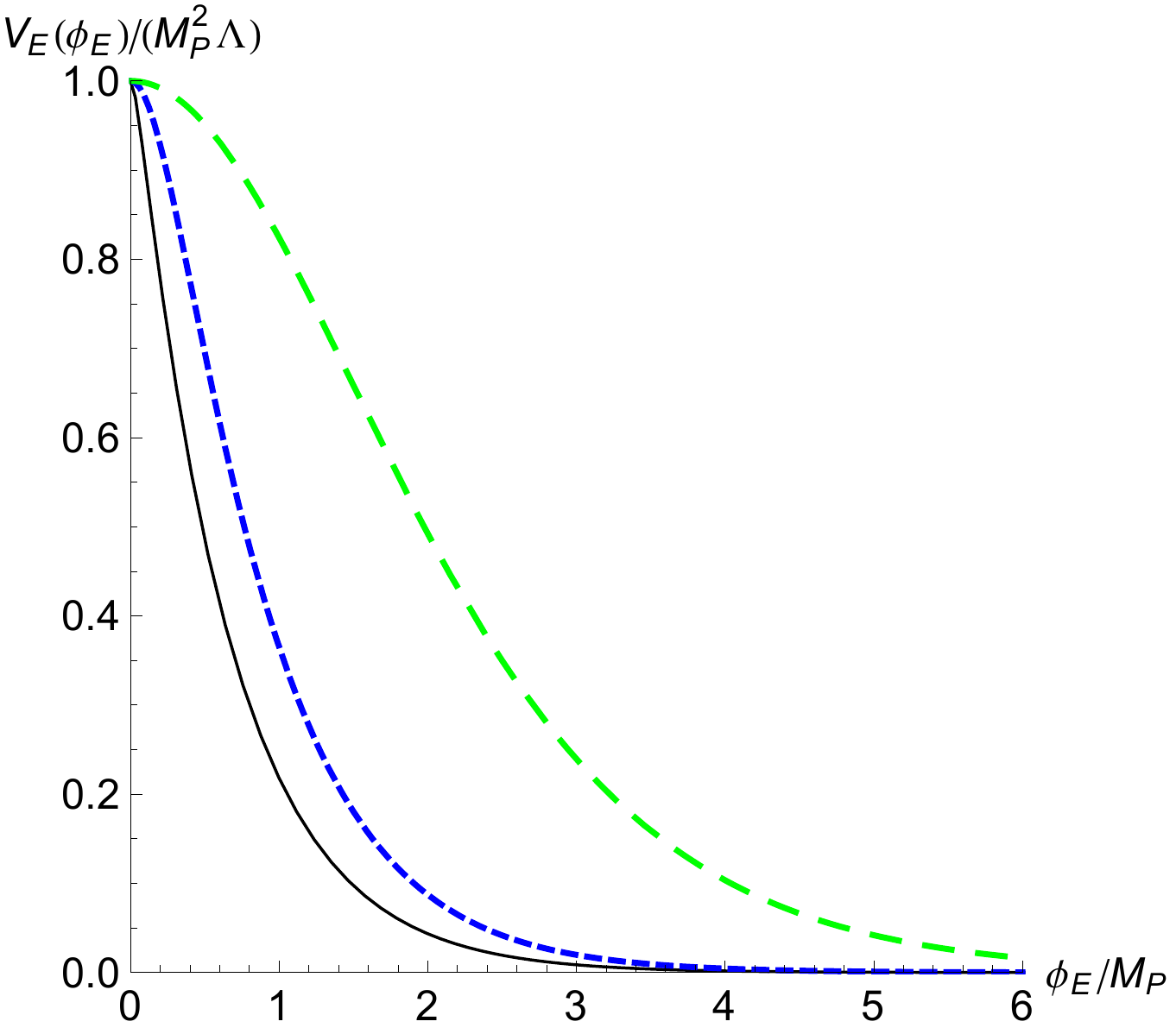}
\caption{The effective potential in the Einstein Frame as a function of the field for $\xi = -10$ (black solid),
 $\xi = -1$ (blue short dashed) and  $\xi = -0.1$ (green long dashed). }
\label{Potential Einstein frame}
\end{figure}

 In this section we discuss the model defined in Eq.~(\ref{action}) in the Einstein frame.  
 To get to the Einstein frame, one ought to perform the following frame (conformal) transformations,
 \begin{equation}
  a_E^2 = \frac{F(\phi)}{M_{\rm P}^2}a^2
  \,,\qquad d\phi_E =\sqrt{ \frac{M_{\rm P}^2}{F(\phi)}\left(1+\frac32\frac{[dF(\phi)/d\phi]^2}{F(\phi)}\right)} \;d\phi
 \,,
 \label{conformal frame transformations}
 \end{equation}
where the index $E$ denotes the Einstein frame. The cosmological constant transforms as the (constant part of the)
corresponding Jordan frame potential,
$V_0=M_{\rm P}^2\Lambda$,
\begin{equation}
   V_E(\phi_E) = \frac{M_{\rm P}^6\Lambda}{F^2(\phi(\phi_E))}
\,.
\label{cosmological constant: Einstein frame}
\end{equation}
This can be easily seen by requiring that $\sqrt{-g}\Lambda$ must be invariant under frame transformations,
from which it follows,  $\sqrt{-g}\Lambda=\sqrt{-g_E}\Lambda M_{\rm P}^4/F^2(\phi)$,
where we made use of $\sqrt{-g}=a^4$. When the above transformations are exacted,
the scalar-tensor action~(\ref{action}) in the Einstein frame becomes simply,
\begin{equation}
  S_E = \int d^4 x\sqrt{-g_E}\left(\frac{M_{\rm P}^2}{2}R_E - \frac12 g_E^{\mu\nu}\partial_\mu\phi_E\partial_\nu\phi_E-V_E(\phi_E)\right)
\,,
\label{action:E}
\end{equation}
where in the case when $V(\phi)=0$, $V_E(\phi_E)$ is given in~(\ref{cosmological constant: Einstein frame}),
making the (effective) cosmological constant field dependent. To get an insight into how
 $V_E(\phi_E)$ depends on $\phi_E$, it is worth devoting some attention to the form of this Einstein frame
 `potential.' Assuming $F(\phi)$ is given by Eq.~(\ref{F and V}),
Eq.~(\ref{conformal frame transformations}) for $\phi_E$ can be integrated, to give,
\begin{eqnarray}
 \frac{\phi_E(\phi)}{M_{\rm P}} = \sqrt{\frac{1\!-\!6\xi}{-\xi}} {\rm Arcsinh}\bigg(\frac{\sqrt{-\xi(1\!-\!6\xi)}\phi}{M_{\rm P}}\bigg)
\label{phiE of phi}
-\sqrt{6} {\rm Arctanh} \Bigg(\frac{\sqrt{6}(-\xi)\phi}{\sqrt{M_{\rm P}^2 -(1\!-\!6 \xi)\xi\phi^2}}\Bigg)
 \,,\;
\end{eqnarray}
where the integration constant is chosen such to get $\phi_E(0)=0$. The Einstein frame potential is shown in
figure~\ref{Potential Einstein frame} (for several values of nonminimal coupling, $\xi = -10,-1,-0.1$).
The potential $V_E$ is a monotonically decreasing function of $\phi_E$.
For small values of the field, $\phi_E\ll M_{\rm P}$, the potential~(\ref{cosmological constant: Einstein frame})
can be approximated by a constant plus a negative mass term,
\begin{equation}
V_E(\phi_E)\simeq \Lambda\left[M_{\rm P}^2+2\xi\phi_E^2\right] +{\cal O}(\phi_E^4),
   \,,
\label{VE small field}
\end{equation}
while for $\phi_E\gg M_{\rm P}$, the potential decays exponentially with the field,
\begin{equation}
V_E(\phi_E)\simeq V_{E0}\exp\left(\!-\!\lambda_E\frac{\phi_E}{M_{\rm P}}\right)
\,,\quad
V_{E0}=16M_{\rm P}^2\Lambda(1\!-\!6\xi)^2 \left[\sqrt{1\!-\!6\xi}\!-\!\sqrt{-\!6\xi} \right]
  ^\frac{4\sqrt{-\!6\xi}}{\sqrt{1\!-\!6\xi}}
  \,.
\label{VE large field}
\end{equation}
where $ \lambda_E = 4\sqrt{-\xi/(1\!-\!6\xi)}$.
The relevant equations of motion in the Einstein frame are,
\begin{eqnarray}
  &&\hskip -1cm
  \ddot \phi_E + 3H_E\dot\phi_E + V_E^\prime(\phi_E) = 0
\nonumber\\
  H_E^2 &=& \frac{1}{3M_{\rm P}^2}\bigg(\frac{\dot\phi_E^2}{2}+V_E(\phi_E)\bigg)
\nonumber\\
\dot H_E &=& -\frac{\dot\phi_E^2}{2M_{\rm P}^2}
\,,
\label{EOM: Einstein frame}
\end{eqnarray}

\begin{figure}[ht]
\begin{minipage}{.45\textwidth}
        \begin{center}
\includegraphics[width=7.5cm,height=7cm]{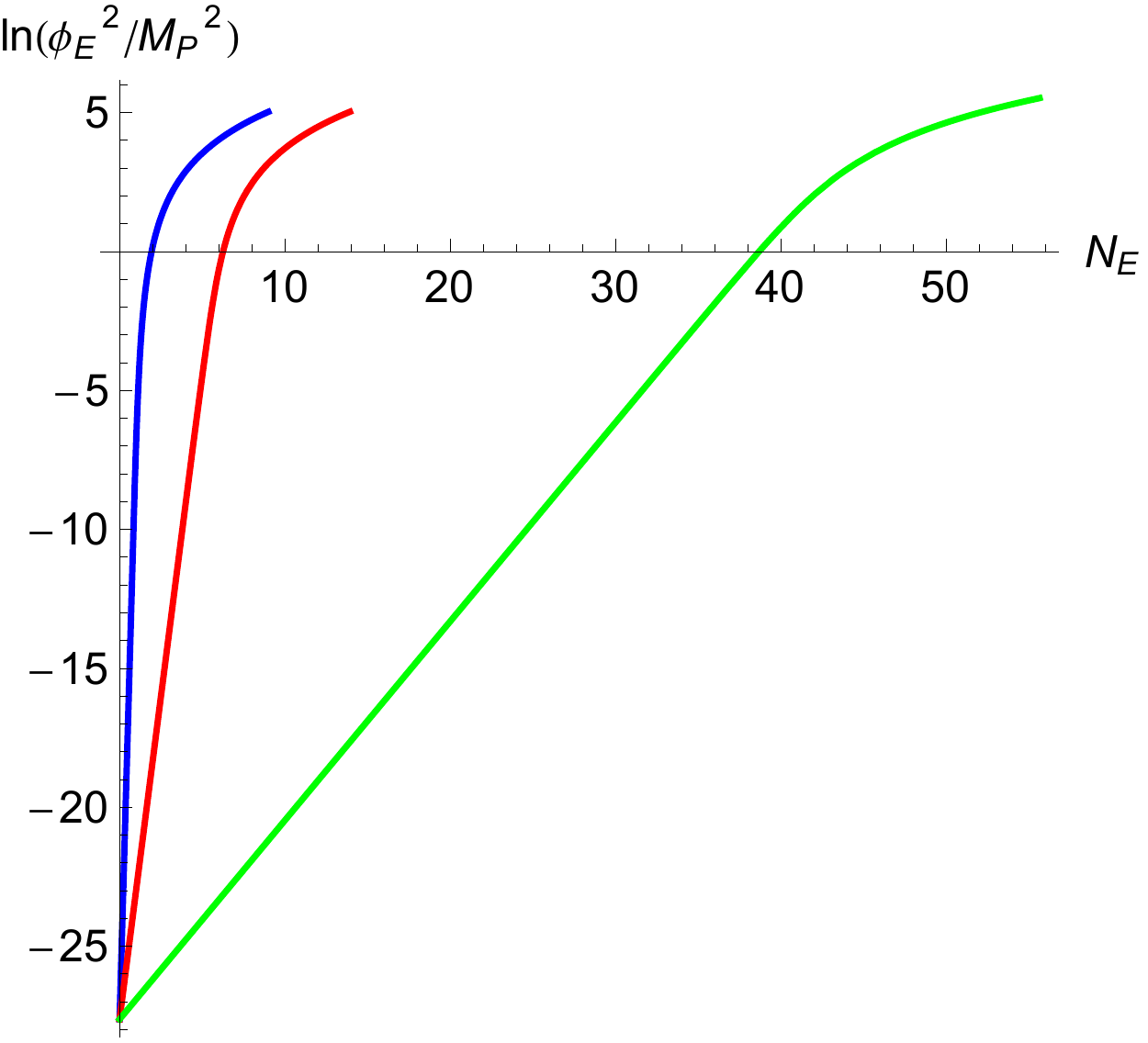}
\end{center}
    \end{minipage}
    \hskip 0.5cm
\begin{minipage}{.45\textwidth}
        \begin{center}
\includegraphics[width=7.5cm,height=7cm]{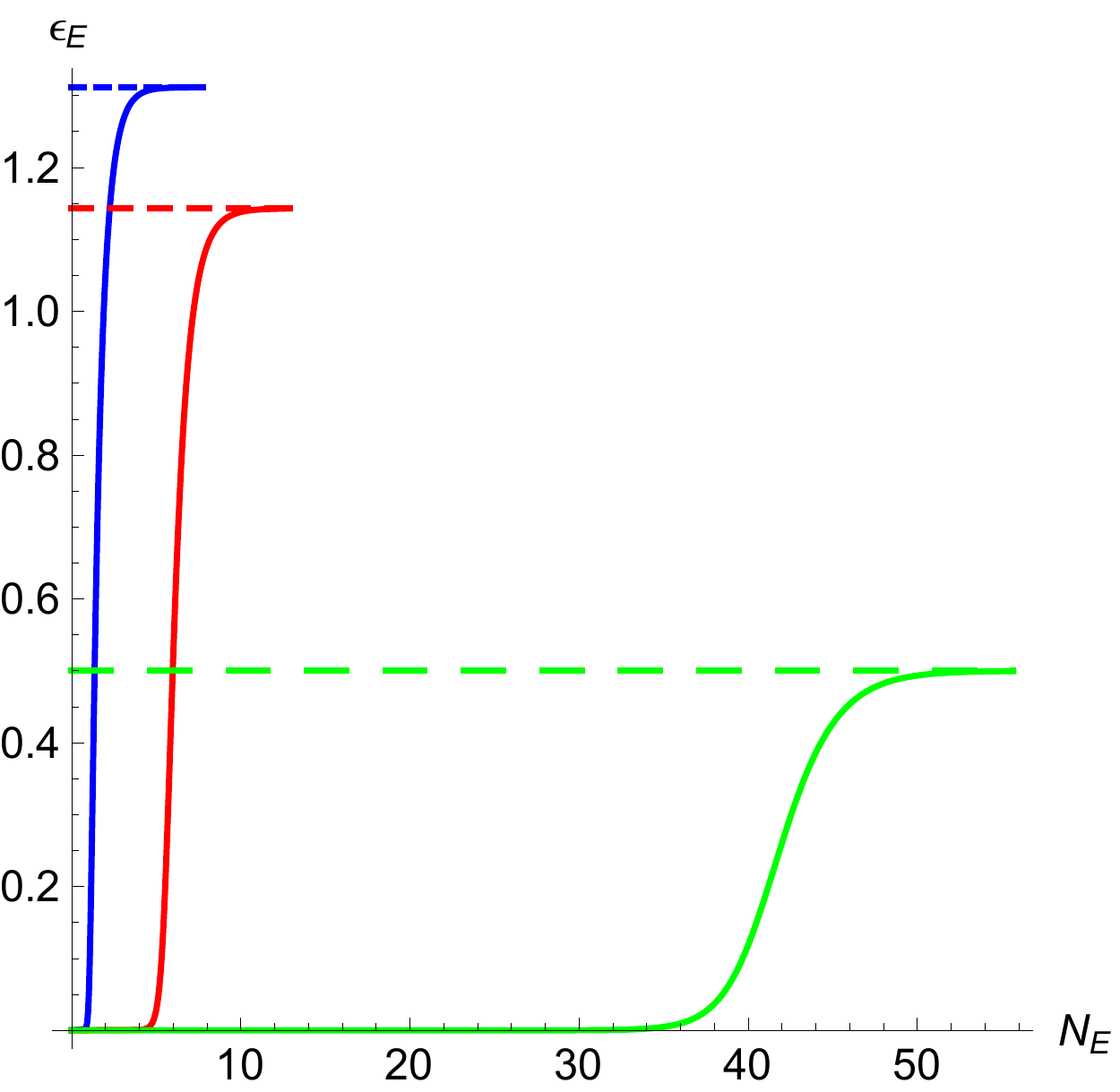}
\end{center}
    \end{minipage}
\caption{{\it Left panel:} $\ln(3\phi_E^2/\Lambda)$ as a function of the number of e-foldings in the Einstein frame.
Just as in the Jordan frame, shown in figure~\ref{Phi2 and epsilon vs N},
both in the quasi-de Sitter stage and in the late power-law expansion stage the field $\phi_E(t)$ grows
exponentially with the number of e-foldings. The curves (from left to right) correspond to
$\xi = -10,-1$ and $\xi=-0.1$, respectively.
{\it Right panel:} The Einstein frame principal slow roll parameter, $\epsilon_E=-\dot H_E/H_E^2$ as a function of
the number of e-foldings. $\epsilon_E$ stays close to zero
during the quasi-de Sitter stage and transits to a constant value $\epsilon_E=-8\xi/(1\!-\!6\xi)$
during the late-time scaling regime. From left to right: $\xi = -10,-1$ and $\xi=-0.1$. In all cases the initial value of the field
is $\phi_0=10^{-6}M_{\rm P}$, which is the typical size of quantum fluctuations during primordial inflation.}
\label{PhiE2 and epsilonE vs NE}
\end{figure}

If the initial field value $\phi_E$ is homogeneous and close to zero, then from Eqs.~(\ref{VE small field}) we see
that the Universe undergoes a relatively brief period of inflation, followed by a period of 
a slow roll parameter $\epsilon_E$ that asymptotes to a constant (see {\it e.g.} Refs.~\cite{Joyce:1997fc,Halliwell:1986ja,Ratra:1987rm},
in which attractors in exponential and power law potentials were considered),
\begin{equation}
  \epsilon_E = \frac{\lambda_E^2}{2}=\frac{-8\xi}{1\!-\!6\xi}
  \,.
\label{epsilonE}
\end{equation}
For a large and negative $\xi$, $\epsilon_E\rightarrow 4/3$, which is a decelerating epoch. The limiting case
(between acceleration and deceleration), $\epsilon_E=1$, is reached when $\xi=-1/2$ 
(this is the same value as in the Jordan frame) in which case the Universe
behaves as if it were spatial curvature dominated. Numerical solutions of Eqs.~(\ref{EOM: Einstein frame})
-- shown in figure~\ref{PhiE2 and epsilonE vs NE} -- confirm this simple picture.
For large and negative $\xi$ and when $\phi_E\gg M_{\rm P}$
the Einstein frame potential $V_E$ induced by a Jordan frame cosmological constant $\Lambda$
behaves as $V_E\propto 1/t^2$, approaching zero at asymptotically late times. From figure~\ref{PhiE2 and epsilonE vs NE} we see that the
Universe enters a late time power law expansion corresponding to the slow roll parameter $\epsilon_E$ given in~(\ref{epsilonE}).
$\epsilon_E$ increases monotonically as $-\xi$ increases, reaching asymptotically $4/3$ as $-\xi\rightarrow \infty$.
Since during radiation and matter eras $\rho_m\propto a_E^{-2\epsilon_m}$, with $\epsilon_m = 2$ and $3/2$, respectively,
the asymptotic scaling $\Lambda_{E\rm eff}\propto a_E^{-2\epsilon_E}\approx  a_E^{-8/3}$ ({\it cf.} Eq.~(\ref{Lambda eff}))
is not enough to solve the cosmological
constant problem. An important question is whether this scaling can be improved by suitably changing
the nonminimal coupling function
$F=F(\phi)$ in Eq.~(\ref{F and V}). We have the following

\begin{quote}
  {\bf  Conjecture:}
  {\it For an arbitrary positive nonminimal coupling function, $F(\phi)>0$, the fastest scaling of the effective cosmological
constant in Einstein frame is $\Lambda_{E\rm eff} \propto a_E^{-8/3}$, {\rm i.e.} $\epsilon_E \leq 4/3$.}
\end{quote}

We are unable to rigorously prove this conjecture. However, we have collected strong evidence
 -- which we summarize in 
Appendix~A --  that supports it.

Therefore, if we want to have a viable solution to the cosmological constant problem, we have to add matter and, as we argue below,
 the field $\phi_E$ must sufficiently quickly decay into matter, which is what we discuss next.


\section{Adding matter fields}
\label{Adding matter fields}

  The simplest way to include matter fields is to add a homogeneous, time dependent perfect fluid, which is
in the fluid rest frame characterized by energy density $\rho_m(t)$, pressure $p_m(t)$ and an equation of state,
which is in cosmological space-times typically of the form, $p_m=w_m\rho_m$, where $w_m$ is a constant
(for a relativistic fluid, $\epsilon_m=2$, for a non-relativistic fluid, $\epsilon_m=3/2$).
In the regime when matter energy density is much smaller than 
the scalar field energy density the interaction of matter with
the TeS gravitational sector can be neglected and the energy density and pressure will scale as,
\begin{equation}
  \rho_m = \frac{\rho_{m0}}{a_E^{2\epsilon_m}}\,, \quad p_m = w_m\frac{\rho_{m0}}{a_E^{2\epsilon_m}}
  \,, \quad \epsilon_m=\frac32(1\!+\!w_m)
  \,,
  \label{classical fluid: scaling}
\end{equation}
where $a_E=a_E(t) = {\rm e}^{N_E(t)}$ is the Einstein frame scale factor and $N_E$ denotes the 
number of e-folds in the Einstein frame.

In presence of matter, Eqs.~(\ref{EOM: Einstein frame}) generalise to
\begin{eqnarray}
  \dot \rho_{\phi E} + 3H_E(\rho_{\phi E}\!+\!p_{\phi E})\! &=&\! -\Gamma(\rho_{\phi E}-g\rho_m)
\,,\qquad \rho_{\phi E}\!+\!p_{\phi E}=\dot\phi_E^2
\label{EOM: Einstein frame with matter:1}\\
  \dot \rho_{m} + 3H_E(\rho_m\!+\!p_m)\! &=&\! \Gamma(\rho_{\phi E}-g\rho_m)
\label{EOM: Einstein frame with matter:2}\\
  H_E^2 &=& \frac{1}{3M_{\rm P}^2}\bigg( \rho_{\phi E}+\rho_m\bigg)
\,,\qquad \rho_{\phi E}=\frac{\dot\phi_E^2}{2}+V_E(\phi_E)
\label{EOM: Einstein frame with matter:3}\\
\dot H_E &=& -\frac{1}{2M_{\rm P}^2}\left(\rho_{\phi E}\!+\!p_{\phi E}+\rho_m+p_m\right)
\,,
\label{EOM: Einstein frame with matter:4}
\end{eqnarray}
where $g=g_\phi/g_m$ (not to be confused with ${\rm det}[g_{\mu\nu}]$)
 is the ratio of the number of relativistic degrees of freedom in the field and in 
matter (it is reasonable to take 
$g_\phi=1$ and $g_m\simeq 100$) and $\Gamma$ is the decay rate at which the field decays into matter.

The rate $\Gamma$ can be a true constant, or it can be time dependent, and its time dependence 
is frame dependent. The details of this frame dependence in some field theoretical models are discussed in Appendix~B;
here we discuss phenomenological models for $\Gamma$ in the Einstein frame. 
Plausible time dependences can be parametrized in terms of 
the Hubble rate as, $\Gamma = \gamma_\theta (H_E/H_0)^\theta H_0$, where $H_0=\sqrt{\Lambda/3}$. 
We shall assume that, at early times (when $H\sim H_0$) $\Gamma\ll H_E$, while at sufficiently late times, $\Gamma\gg H_E$, which 
can be realised when $0\leq \theta\leq 1$. In this paper we consider in detail two cases:
\noindent
\begin{quote}
 {\bf Case A:} $\Gamma=\gamma_0 H_0={\rm constant}$, and \\
{\bf Case B:} $\Gamma=\gamma_1 H_E$, with $\gamma_1={\cal O}(1)$. 
\end{quote}
Below we  argue that late time results for the more general case when $0<\theta<1$ can be subsumed in Case~A.

Assuming that initially $\rho_m\ll \Lambda M_{\rm P}^2$ (which is typically the case in the early Universe setting, even during
phase transitions induced by a Higgs-like field), from Eqs.~(\ref{EOM: Einstein frame with matter:1}--\ref{EOM: Einstein frame with matter:4}) 
one can study the time dependence of
the Hubble parameter $H_E$ and the corresponding principal slow roll parameter $\epsilon_E$.

 {\bf Case A:} $\Gamma=\gamma_0 H_0$, such that at late times, $\Gamma\gg H$ (tight coupling limit), 
which is implied by the scaling of the Hubble parameter: 
it scales at least as $H\propto 1/t$. In this case, at sufficiently late times, $\Gamma\gg H_E$ enforces, 
\begin{equation}
  \rho_{\phi E}=g\rho_m
\,,
\label{locking of rhos}
\end{equation}
plus small corrections. The condition~(\ref{locking of rhos}) becomes exact in thermal equilibrium.
Note that Eq.~(\ref{locking of rhos}) simplifies Eq.~(\ref{EOM: Einstein frame with matter:2}) to 
\begin{equation}
  \dot \rho_{m} + 2\epsilon_m H_E\rho_m \simeq 0\,,\qquad \epsilon_m = \frac32(1\!+\!w_m)
\,,
\label{EOM: Einstein frame with matter:2b}
\end{equation}
which is solved by, $\rho_m=\rho_{m0}/a_E^{2\epsilon_m}$. The condition~(\ref{locking of rhos})) 
implies that the scalar field energy scales the same way, {\it i.e.}
$\rho_{\phi E}=\rho_{\phi E 0}/a_E^{2\epsilon_\phi}$, with 
\begin{equation}
  \epsilon_{\phi} = \epsilon_m
\,,\qquad \epsilon_{\phi} = \frac32\frac{\rho_{\phi E}+p_{\phi E}}{\rho_{\phi E}}= \frac32\frac{\dot\phi_E^2}{\rho_{\phi E}}
\,,
\label{consistency on epsilons}
\end{equation}
and $\rho_{\phi E 0}=g \rho_{m0}$.
The condition~(\ref{consistency on epsilons}) must be consistent with the solution of the Friedmann 
equations~(\ref{EOM: Einstein frame with matter:3}-\ref{EOM: Einstein frame with matter:4}), from which we extract, 
\begin{eqnarray}
 \epsilon_E &=& -\frac{\dot H_E}{H_E^2} = \frac{3}{2}\frac{\dot\phi_E^2+\frac23\epsilon_m\rho_m}{\rho_{\phi E}+\rho_m}
   = \frac{g\epsilon_\phi + \epsilon_m}{g+1} = \epsilon_m
\,,
\end{eqnarray}
which is consistent with~(\ref{consistency on epsilons}). We have thus proved that 
\begin{equation}
  \rho_{\phi E} =g\rho_m=g \frac{\rho_{m0}}{a_E^{2\epsilon_m}}
\,;\qquad p_{\phi E}=\Big(\frac23\epsilon_m \!-\!1\Big)\rho_{\phi E}
\,,\quad p_{m}=\Big(\frac23\epsilon_m \!-\!1\Big)\rho_{m}
\,.
\label{solution Gamma gg HE}
\end{equation}
We think that this solution is the late time attractor, but we were unable to prove it (an attempt to numerically solve
Eqs.~(\ref{EOM: Einstein frame with matter:1}--\ref{EOM: Einstein frame with matter:4}) failed because these
equations become stiff at late times).

{\bf Case B:} $\Gamma=\gamma_1 H_E$, with $\gamma_1={\cal O}(1)$. Making the
constant late time $\epsilon_E$ {\it Ans\"atze}, 
\begin{equation}
\rho_m=\frac{\rho_{m0}}{a_E^{2\epsilon_E}}
\,,\qquad \rho_{\phi E}=\frac{\rho_{\phi E0}}{a_E^{2\epsilon_E}}
\,,\qquad  r =\frac{\rho_{\phi E}}{\rho_m} = {\rm const.}
\label{Anasatz case B}
\end{equation}
and inserting them into  Eqs.~(\ref{EOM: Einstein frame with matter:1}--\ref{EOM: Einstein frame with matter:2}) yields, 
\begin{eqnarray}
 -\epsilon_E +\epsilon_\phi \! &=&\!  \frac{\gamma_1}{2r}(g\!-\!r)
\label{EOM: Einstein frame with matter:1b}\\
  \delta\epsilon\equiv \epsilon_E-\epsilon_m\! &=&\!  \frac{\gamma_1}{2}(g\!-\!r)
\label{EOM: Einstein frame with matter:2b}
\end{eqnarray}
where $\epsilon_\phi = (3/2)(\dot\phi_E^2/\rho_{\phi E})$. Next, by combining 
Eqs.~(\ref{EOM: Einstein frame with matter:3}--\ref{EOM: Einstein frame with matter:4}) one finds, 
\begin{eqnarray}
\frac{V_{E0}t_0^2}{M_{\rm P}^2} &=& \frac{3}{\epsilon_E^2}- \frac{2}{\lambda_E^2}
                        - \frac{3}{\epsilon_E\epsilon_m}+ \frac{\epsilon_E}{\epsilon_m\lambda_E^2}
\,,\quad \frac{\rho_mt^2}{M_{\rm P} 2} = \frac{3}{\epsilon_m}\left(\frac{1}{\epsilon_E}- \frac{2}{\lambda_E^2}\right) 
\label{EOM: Einstein frame with matter:3b}\\
\epsilon_\phi &=& \frac{\epsilon_m}{1+ \frac{\lambda_E^2}{2}\frac{\epsilon_E-\epsilon_m}{\epsilon_E^2}}
\,.
\label{EOM: Einstein frame with matter:4b}
\end{eqnarray}
Inserting the latter equation into Eq.~(\ref{EOM: Einstein frame with matter:1b}) results in the following quartic equation for
$ \delta\epsilon\equiv \epsilon_E-\epsilon_m=(\gamma_1/2)(g\!-\!r)$,
\begin{eqnarray}
 &&\delta\epsilon\bigg\{\delta\epsilon^3 + \bigg[2\epsilon_m-\frac{\lambda_E^2}{2}-\frac{\gamma_1}{2}(g\!+\!1)\bigg] \delta\epsilon^2
\label{quartic equation}\\
   &&\hskip 1cm
+\,\bigg[\epsilon_m\Big(\epsilon_m-\frac{\lambda_E^2}{2}\Big)-\frac{\gamma_1}{2}(g\!+\!1)\Big(2\epsilon_m-\frac{\lambda_E^2}{2}\Big)\bigg] \delta\epsilon
 -\frac{\gamma_1}{2}\epsilon_m\bigg[\epsilon_m+g\Big(\epsilon_m-\frac{\lambda_E^2}{2}\Big)\bigg]\bigg\}=0
\,.
\nonumber
\end{eqnarray}
\begin{figure}[ht]
\begin{minipage}{.45\textwidth}
        \begin{center}
\includegraphics[width=7.5cm,height=7cm]{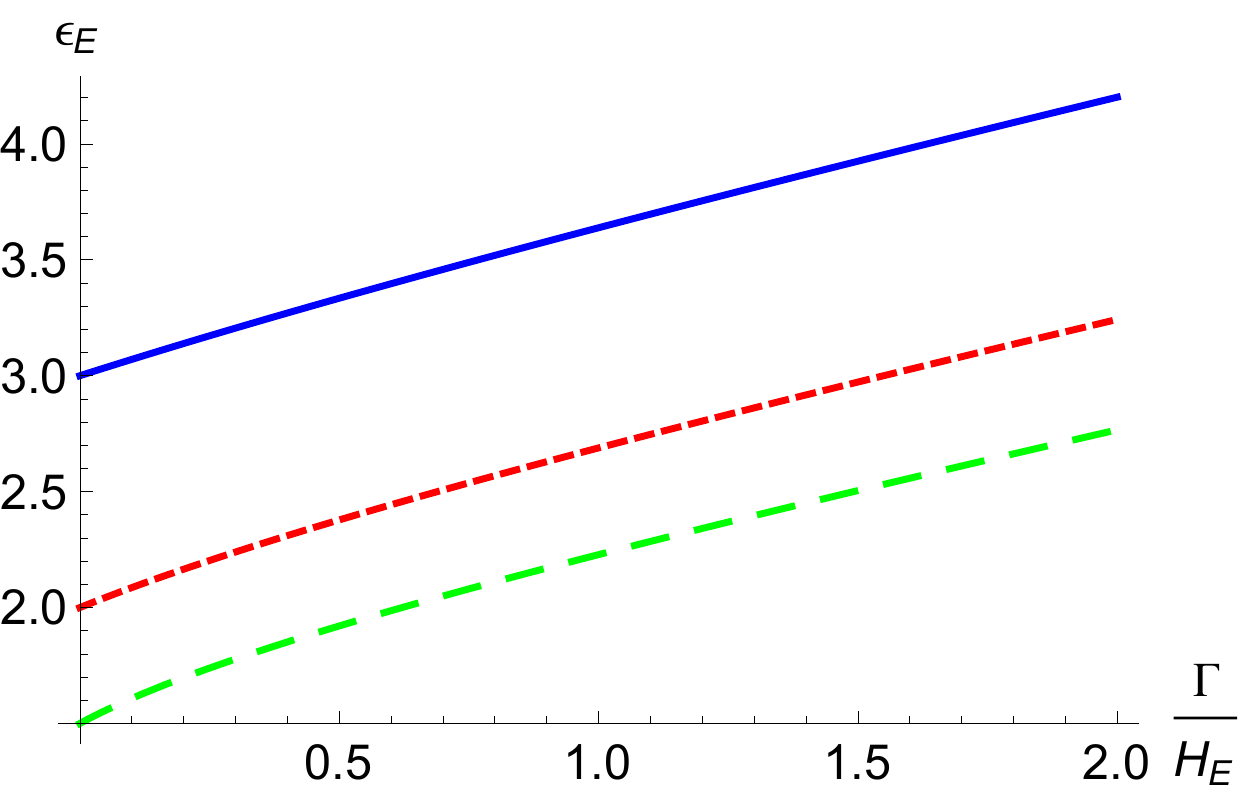}
\end{center}
    \end{minipage}
    \hskip 0.5cm
\begin{minipage}{.45\textwidth}
        \begin{center}
\includegraphics[width=7.5cm,height=7cm]{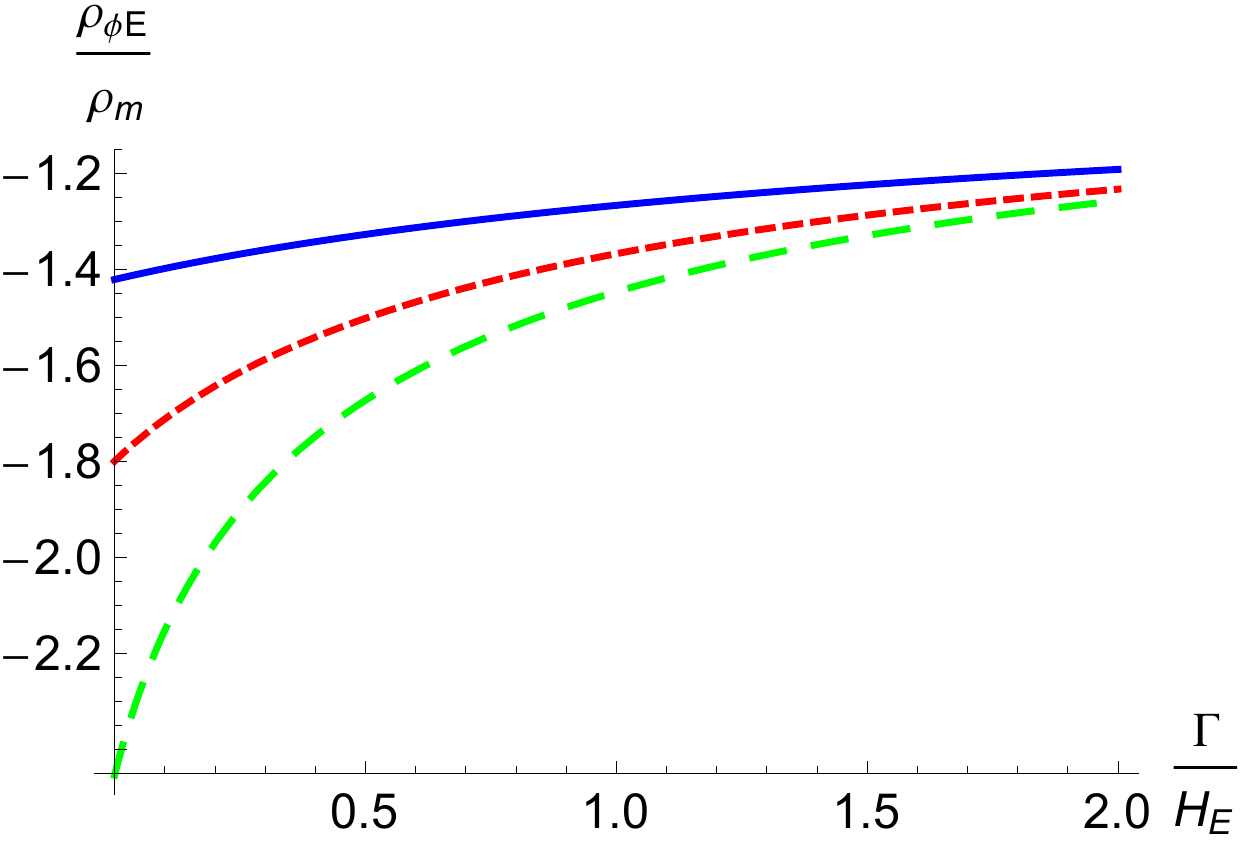}
\end{center}
    \end{minipage}
\caption{{\it Left panel:}  $\epsilon_E$ for the real solution of the cubic equation~(\ref{quartic equation})
as a function of the decay strength $\Gamma/H_E$.
Note that $\epsilon_E\geq \epsilon_m$ and that it grows approximately linearly with $\Gamma/H_E$.
The curves (from top to down) correspond to
$\epsilon_m =3$ (solid blue),  $\epsilon_m =2$ (dashed red), and $\epsilon_m =3/2$ (green, long dashed), respectively.
{\it Right panel:} $r=\rho_{\phi E}/\rho_m$ for the real solution of the cubic equation~(\ref{quartic equation})
as a function of the decay strength $\Gamma/H_E$.
Note that $r$ is negative. 
 The curves (from top to down) correspond to
$\epsilon_m =3$ (solid blue),  $\epsilon_m =2$ (dashed red), and $\epsilon_m =3/2$  (green, long dashed), respectively.
In all cases the parameters chosen are $\lambda_E=4/3$ and  $g = 1/10$.
}
\label{epsilonE and r vs Gamma}
\end{figure}
This equation has two real solutions and two complex solutions, which we immediately discard.
The first real solution is $\epsilon_E=\epsilon_m$, $r=g$, and the second is plotted in figure~\ref{epsilonE and r vs Gamma}
for typical choices of the parameters. 
Numerical investigation of these solutions shows that, for positive $\Gamma$,  $\epsilon_E>\epsilon_m$ and $r=\rho_{\phi E}/\rho_m<0$
(see  figure~\ref{epsilonE and r vs Gamma}),
rendering the second solution unphysical. The first solution is also unphysical because $\epsilon_E=\epsilon_m>\lambda_E^2/2$ 
is only possible if $\rho_m<0$, {\it cf.} Eq.~ (\ref{EOM: Einstein frame with matter:3b}). This forces us to conclude that,
in the case when $\Gamma=\gamma_1 H_E$,  there are no asymptotic solutions for which both $\rho_m$ and $\rho_{\phi E}$ are 
positive and $\rho_{\phi E}/\rho_m={\rm constant}$.   

 We have thus found out that, to get an effective decay of $V_E$, $\Gamma=\gamma_\theta(H_E/H_0)^\theta$ 
has to grow with respect to $H_E$, {\it i.e} $\theta<1$. 

In Appendix~B we consider some simple (tree-level) decay channels, in which the fields couple canonically in the 
Jordan frame. These include a Yukawa interaction, a fermionic mass term, a second scalar ($\chi$-)field mass term and 
a cubic (and quartic) term of the type, $\sim (m_\chi^2+\sigma \phi)\chi^2$ and a canonical coupling to photons.  
Surprisingly, we find that (at tree level) none of them produces a viable decay channel of the type considered in Case~A
above. The processes that work are a non-canonical coupling of  photons to the gravitational scalar $\phi$ (examples of $G(\phi)$ 
in Eq.~(\ref{typical interactions}) that work are $G(\phi)\propto 1/\phi^2$ and $G(\phi)\propto 1/\phi^4$)
and a decay of $\phi$ into a massless scalar $\chi$ (induced by a bilinear coupling term $\propto\phi\chi$).

\section{Discussion and conclusions}
\label{Discussion and conclusions}  

 We have constructed a simple model of tensor-scalar gravity  in which the cosmological constant generated in the Jordan frame 
decays sufficiently fast to be suppressed during radiation and matter eras. 
In order for this to represent a satisfactory solution to the cosmological constant problem,
one must not mess up with any of the processes in the early universe that are known to work. These include: 
nucleosynthesis, photon decoupling and the growth of large scale structure
(not enough is known about inflation, baryogenesis, dark matter production, neutrino decoupling, and 
cosmic neutrino background to place meaningful constraints on our model).

 Nucleosynthesis and photon decoupling remains unaffected by the scenario considered here, because 
in the tight coupling limit discussed in section~\ref{Adding matter fields}, Case A, 
both the effective cosmological constant 
$V_E$ and matter density scale in the Einstein frame as free matter does; for example $\epsilon_E=2$ and 
$\epsilon_E=3/2$ for relativistic and nonrelativistic fluids, respectively.

 The tight coupling regime in section~\ref{Adding matter fields} means that the cosmological constant contribution 
$V_E$ and matter $\rho_m$ remain in balance  throughout the history of the Universe (from the moment
the tight coupling approximation holds). During that time the scalar field $\phi$ incessantly decays into matter.
The decay channel must remain open up to late times, which means that $\phi$ must decay (not only) into 
baryons and leptons of the standard model, but also to (almost) massless particles such as photons and 
massless scalars (which must be therefore chosen outside the standard model). If one produces primarily 
photons, their energy will redshift fine (during matter era), but one will be able to see this as 
a faint diffuse background source of very long wavelength stochastic photon radiation which, when included,
will increase the photon-to-baryon ratio, leading thus to a potentially observable phenomena.
If the decaying product is an ultralight scalar outside the standard model, its energy density will redshift as 
radiation, and at the moment we do not see how one would observe it.

As regards the growth of structure, since background matter evolves according to the standard law, 
$\rho_m\propto 1/a^{2\epsilon_m}$, where $2\epsilon_m=3(1\!+\!w_m)$ and $w_m=p_m/\rho_m$, 
matter perturbations will perceive the Einstein frame Hubble rate, $H_E=1/(\epsilon_m t)$, 
and thus grow in the same way as in the standard (minimally coupled) cosmology.
At late times (after nucleosynthesis, at which $H_E\sim 10^{-17}~{\rm eV}$) the production of 
electrons, baryons or even neutrinos is kinematically forbidden,
and hence the effective cosmological constant decays cannot affect the growth of structure. These remarks imply 
that in our model we expect the same growth of structure as in the standard $\Lambda$CDM model. 

Note that our model does not violate the Weinberg's no go theorem~\cite{Weinberg:1988cp}
for the following reasons. Strictly speaking, Weinberg's theorem applies to static situations, in which one seeks a solution
to the cosmological constant by adjusting the scalar field value to a suitable constant, 
and the corresponding metric is also time independent, while in our model both  
the scalar field and the metric are dynamical. 
Further criticism of the scalar field adjustment mechanisms in~\cite{Weinberg:1988cp}
refers to the observation that any adjustment mechanism 
implies in the Jordan frame an effective Newton constant that asymptotically vanishes in time. 
We address this issue by postulating that the observers (us) perceive the Universe expansion 
 in the Einstein frame, in which the effective Newton constant does 
not change in time. 

 At the end we briefly discuss possible dark energy candidates.
One possibility is the contribution from (light, minimally or non-minimally coupled scalar field) inflationary 
fluctuations; the details are discussed in 
Refs.~\cite{Glavan:2013mra,Glavan:2014uga,Glavan:2015,Ringeval:2010hf,Aoki:2014dqa} and 
the contribution from neutrinos and photons (assuming they couple minimally in the Einstein frame such that their 
contribution remains constant in the physical Einstein frame). Other proposals involve modified gravity or 
scalar field (quintessence) models, see {\it e.g.} Ref.~\cite{Copeland:2006wr}.

 Here we will briefly consider another possibility. Since neutrino masses cannot be explained within the standard model,
it is quite natural to postulate that they couple minimally in the Einstein frame. If so, 
their contribution to the cosmological
constant does not scale away with time.~\footnote{This contribution comes from the virtual vacuum fluctuations which 
dominates at late times, 
and should not be confused with the classical contribution which scales as $\propto 1/a^3$ at late times,
and thence becomes eventually subdominant.} 
Let us calculate it. 
Imposing a Lorentz invariant regularization, the cosmological constant produced by three light Majorana neutrinos
is (the fermionic contribution per relativistic degree of freedom is the same as that of a massive scalar field in Eq.~(11) of Ref.~\cite{Koksma:2011cq}, but with an opposite sign),
\begin{equation}
  \langle \hat \rho_\nu^{\rm ren}\rangle = - \langle \hat p_\nu^{\rm ren}\rangle 
   = - 2\sum_{i=1}^3\frac{m_i^4}{64\pi^2} \ln\left(\frac{m_i^2}{\mu^2}\right) + C(\mu)
\,,
\label{neutrinos to cc}
\end{equation}
(here $C(\mu)$ is an arbitrary constant to be fixed by measurements and which runs logarithmically with $\mu$) 
while the photons do not contribute because they are massless 
(the contributions proportional to the curvature invariants squared are small and can be neglected at late times). 
While we do not know what are neutrino masses, from the results of the MINOS experiment~\cite{Adamson:2011ig}
we know that 
the neutrino mass difference squared between the second and third generation of neutrinos is about 
$|\Delta m_{23}|^2\sim 0.0023~{\rm eV}^2$, which implies that at least one of the neutrinos
is $m_i\sim 4\times 10^{-11}~{\rm GeV}$. Inserting this into~(\ref{neutrinos to cc}) gives 
\begin{equation}
\langle \hat \rho_\nu^{\rm ren}\rangle\sim -8\times 10^{-45}~{\rm GeV}^4\times  
   \ln\left(\frac{2\times 10^{-21}~{\rm GeV}^2}{\mu^2}\right) + C(\mu)
\,.
\label{neutrinos to cc:2}
\end{equation}
This needs to be compared with the dark energy density today, $\rho_{\rm DE}\sim 2\times 10^{-47}~{\rm GeV}^4$,
which means that one ought to tune $C(\mu)$ in~(\ref{neutrinos to cc:2}) at a $1\%$ level
if the neutrino contribution is to explain the dark energy today.

\section*{Acknowledgements}

This work is part of the D-ITP
consortium, a program of the Netherlands Organization for Scientific
Research (NWO) that is funded by the Dutch Ministry of Education,
Culture and Science (OCW).

\section*{Appendix A: Evidence for the Conjecture in section III}
\label{Appendix A: Evidence for the Conjecture in section III}

In section~\ref{Einstein frame analysis} we stated the following 

\begin{quote}
  {\bf  Conjecture:}
  {\it For an arbitrary positive nonminimal coupling function, $F(\phi)>0$, the fastest scaling of the effective cosmological
constant in the Einstein frame is $\Lambda_{E\rm eff} \propto a_E^{-8/3}$, {\rm i.e.} $\epsilon_E \leq 4/3$.}
\end{quote}

Here we present evidence that supports it.

It is instructive to divide $F$ in three classes: (A) $F$'s that asymptotically (for large $\phi$)
grow faster than $\phi^2$; (B) $F$'s that grow slower than
$\phi^2$ (the limiting case, $F\propto \phi^2$, has already been discussed
in section~\ref{Einstein frame analysis}) 
and (C) $F$'s that decrease asymptotically as $\phi$ increases. 
Below we discuss the limit $\phi\rightarrow \infty$ because we are interested in late time behavior. 

In {\bf Class A}, ${F^\prime}^2\gg F$ and
Eq.~(\ref{conformal frame transformations}) can be integrated to give,
\begin{equation}
 F(\phi) \simeq \sqrt{\frac{2}{3}} \phi_E M_{\rm P}
 \,,
\label{phiE case A}
\end{equation}
such that,
\begin{equation}
   V_E  \simeq  \frac{\Lambda M_P^4}{\phi_E^2}\,.
   \label{VE case A}
\end{equation}
In this case asymptotically $\epsilon_E\rightarrow t^{-2/3}\rightarrow 0$ 
as $\phi_E\rightarrow \infty$, and one gets asymptotically de Sitter space.

 {\bf Class B} is more difficult to prove. Here $F\gg {F^\prime}^2$, in which 
case~(\ref{conformal frame transformations}) reduces to,
\begin{equation}
 \phi_E = \int \frac{d\phi}{\sqrt{F(\phi)}}
 \,.
\label{phiE case B}
\end{equation}
This cannot be solved for general $F$. Let us therefore consider the following simple case, $F\propto \phi^{2\omega}$, where $0<\omega<1$.
In this case~(\ref{phiE case B}) can be integrated to give, $\phi_E\propto \phi^{1-\omega}$, $F\propto \phi_E^{2\omega/(1-\omega)}$ such
that
\begin{equation}
 V_E \propto \phi_E^{-4\omega/(1-\omega)}
 \,,
\label{phiE case B}
\end{equation}
which is a negative power potential. We know that a negative power potential, $V_E\propto \phi_E^{-n}$ ($n>0$) 
gives at asymptotically late times, $\epsilon_E(t) \propto t^{-4/(n+4)}=t^{-(1-\omega)}\rightarrow 0$.
In the special case, when $n=0$ ($\omega=0$),  one gets a logarithmic potential, $F\propto \ln(\phi_E/M_{\rm P})$, 
in which case $\epsilon\propto 1/t\rightarrow 0$.

 {\bf Class C} problems can be analysed by considering $F$'s that take the following form, 
$F\propto \phi^{2\omega}$ (for large $\phi$ and $\omega<0$).
As above, this then implies $V_E\propto \phi_E^{-4\omega/(1-\omega)}$. One might hope that, when $\omega\rightarrow -\infty$, 
$V_E\propto \phi_E^4$ and one could get $\epsilon_E\rightarrow 2$. However, the hope is shattered when one realises that
the condition, ${F^\prime}^2\ll F$ implies $\phi,\phi_E\gg M_{\rm P}$ and in that regime the quartic potential 
gives an inflationary slow roll parameter, $0<\epsilon_E<1$. 

 To gain more confidence that our conjecture holds, it is instructive to consider the inverse problem. 
Namely, we know that the Einstein frame potentials,
(1) $V_E(\phi_E) = V_{E0}\exp(-\lambda_E\phi_E/M_{\rm P})$ with $\lambda_E\geq\sqrt{4}$ and
$V_E(\phi_E)=\lambda_n\phi_E^n/M_{\rm P}^{n-4}$ with $n\geq 4$ could be used to get a sufficiently fast scaling of $V_E$ with time to
suppress the effective Einstein frame cosmological constant. Indeed, in the former case we have $\epsilon_E \rightarrow  \lambda_E^2/2\geq 2$
and in the latter case the energy density averaged over an oscillation cycle will scale as, $\epsilon_E\simeq 3n/(n+2)$~\cite{Joyce:1997fc},
such that $\epsilon_E\geq 2$ when $n\geq 4$.
This of course will be true if the field starts at a large (super-Planckian) value, and ends up oscillating around the origin.

 The inverse reconstruction problem can be exacted by firstly observing that Eq.~(\ref{conformal frame transformations}) can be rewritten as,
 \begin{equation}
 d\phi_E \sqrt{F-\frac32\frac{[dF/d\phi_E]^2}{F}}=M_{\rm P}d\phi
 \,.
 \label{conformal frame transformations:reconstruction}
 \end{equation}
In the first case, $F=F_0 \exp[(\lambda_E/2)\phi_E/M_{\rm P}]$ 
and Eq.~(\ref{conformal frame transformations:reconstruction})
reduces to $d\phi_E \sqrt{F(1-3\lambda_E^2/8)}=M_{\rm P}d\phi$ which then implies,
\begin{equation}
   F(\phi) = \frac{\lambda_E^2}{8-3\lambda_E^2}\phi^2\,,\quad \lambda_E^2 < \frac83 \;\Leftrightarrow\; \epsilon_E < \frac43
\,.
\label{F as function of phi}
\end{equation}
The first inequality comes from the requirement $F > 0$ (which must be the case if we demand attractive gravity, {\it i.e.} a positive
effective Newton constant), which then turns into the condition $\epsilon_E <\frac43$, 
which does not give a  fast enough scaling of the effective cosmological constant. 
One can try to repair the problem by adding a constant to $F$ and still keeping the coefficient of the $\phi^2$ term negative. This does not work, 
because that would change the scaling for small field value such that one would enter an accelerating stage with $\epsilon_E<1$.

 In the second case, $V_E(\phi_E)=\lambda_n\phi_E^n/M_{\rm P}^{n-4}$ and $\epsilon_E\rightarrow 3n/(n\!+\!2)$.
The corresponding equation for $F$ is 
 \begin{equation}
 d\phi_E \sqrt{F\bigg(1-\frac{3n^2}{8}\frac{M_{\rm P}^2}{\phi_E^2}\bigg)}=M_{\rm P}d\phi
 \,,
 \label{conformal frame transformations:reconstruction:2}
 \end{equation}
which admits real solutions only when $\phi_E>M_{\rm P}\sqrt{3n^2/8}$, in which the potential $V_E$ drives an accelerating expansion with 
$0<\epsilon_E<1$.  

All of these cases present a sufficient evidence to support the above conjecture.

\section*{Appendix B: Decay rate estimates for simple tree-level decay channels}
\label{Appendix B: Decay rate estimates for simple tree-level decay channels}

In this Appendix we consider some simple decay channel candidates, to see if one can get a growing $\Gamma/H_E$.
Arguably, the simplest decay channels are the ones included by a Yukawa coupling to fermions $\psi$, 
a cubic coupling to some scalar $\chi$ and a (non-)canonical coupling to photons $A_\mu$. 
The corresponding interaction Lagrangian (in the Jordan frame) is,
\begin{equation}
  \sqrt{-g}{\cal L}_{\rm int}=
    \sqrt{-g}\bigg(-(m_\psi+y\phi)\bar \psi \psi -\frac12(m_\chi^2+\sigma\phi)\chi^2 - \frac14 G(\phi)F_{\mu\nu}F^{\mu\nu}\bigg)
\,,
\label{typical interactions}
\end{equation}
where $y$ denotes a Yukawa coupling,  $m_\psi$ a fermion mass, $m_\chi$ is a mass of $\chi$, 
$\sigma$ is a cubic coupling to $\chi$ 
and $G(\phi)=1$ in the case of a canonical coupling to photons. To transform this Lagrangian into the Einstein frame, recall that the fields 
transform as, $\chi_E = \chi/\sqrt{F(\phi)/M_{\rm P}^2}$, 
$A_\mu^E=A_\mu$ and $\psi_E=\psi/[F(\phi)/M_{\rm P}^2]^{3/4}$, such that we get in the Einstein frame,
\begin{equation}
\sqrt{-g_E}{\cal L}_{\rm int E}=
       \sqrt{-g_E}\left(-\frac{m_\psi\!+\!y\phi(\phi_E)}{\sqrt{F/M_{\rm P}^2}}\bar \psi_E \psi_E  -\frac{1}{2}\frac{m_\chi^2\!+\!\sigma\phi(\phi_E)}{F/M_{\rm P}^2}\chi_E^2
                   - \frac14 G(\phi(\phi_E))F^E_{\mu\nu}F_E^{\mu\nu}\right)
\,,
\label{typical interactions: Einstein frame}
\end{equation}
which, in the asymptotic regime when $\phi\gg M_{\rm P}$, simplifies to, 
\begin{eqnarray}
{\cal L}_{\rm int E}&\rightarrow&
     -\Big(m_\psi{\rm e}^{-\lambda_E\phi_E/4M_{\rm P}}\!+\!\tilde yM_{\rm P}\Big)\bar \psi_E \psi_E  
-\frac{1}{2}\bigg(\tilde m_\chi^2{\rm e}^{-\lambda_E\phi_E/2M_{\rm P}}\!+\!\tilde\sigma M_{\rm P}{\rm e}^{-\lambda_E\phi_E/4M_{\rm P}}\bigg)\chi_E^2
\nonumber\\
                  && -\, \frac14 G(\phi(\phi_E))F^E_{\mu\nu}F_E^{\mu\nu}
\,.
\label{typical interactions: Einstein frame:2}
\end{eqnarray}
The corresponding tree level decay rates for the processes $\phi_E\rightarrow 2\; {\rm particles}$ are given by 
\begin{equation}
    \Gamma\sim  \frac{y_{\rm eff}^2}{4\pi}\omega_E + \frac{\sigma_{\rm eff}^2}{4\pi\omega_E}
\,,
\label{tree decays}
\end{equation}
where $y_{\rm eff}$ and $\sigma_{\rm eff}$ are the effective Yukawa and cubic couplings, respectively, and $\omega_E$  is 
the energy of $\phi_E$ excitations, which is for the above case given by, 
\begin{equation}
 \omega_E^2=\frac{d^2V_E}{d\phi_E^2}\simeq \frac{\lambda_E^2 V_E}{M_{\rm P}^2}\sim H_E^2
\,.
\label{omega E}
\end{equation}
From~(\ref{typical interactions: Einstein frame:2}) we see that, $y_{\rm eff}\sim (m_\psi/M_{\rm P})(H_E/H_0)^{1/2}$
and $\sigma_{\rm eff}\sim (m_\chi^2/M_{\rm P})(H_E/H_0)+\tilde\sigma(H_E/H_0)^{1/2}$ (the contribution from 
the photons vanishes when $G=1$). Thence, we get the following estimate for the decay rate~(\ref{tree decays}),
\begin{equation}
 \Gamma\sim \frac{m_\psi^2}{M_{\rm P}^2}\frac{H_E^2}{H_0}+  \frac{m_\chi^4}{M_{\rm P}^2}\frac{H_E}{H_0^2}
                              + \frac{\tilde \sigma^2}{H_0}
\,.
\label{tree decays:2}
\end{equation}
This formula implies that only the last contribution grows with respect to the Hubble rate, and hence it is a potential candidate for the 
decay channel. For this decay to be kinematically allowed, $\omega_E> 2 m_{\chi\rm eff}$, where 
$m_{\chi\rm eff}\simeq\sqrt{\tilde \sigma M_{\rm P}}(H_E/H_0)^{1/4}$ 
is the effective $\chi$ mass that corresponds to that decay channel. Thus we have the following conditions,
\begin{equation}
  \frac{\tilde \sigma^2}{H_0} \gg H_E
\,,\qquad H_E \gg \sqrt{\tilde \sigma M_{\rm P}}\Big(\frac{H_E}{H_0}\Big)^{1/4}
 \;\Longrightarrow\; \frac{H_E}{H_0}\gg \Big(\frac{ M_{\rm P}}{H_0}\Big)^2
\,.
\label{tree decays:3}
\end{equation}
which cannot be satisfied since $H_E\ll H_0$ (due to the evolution) and $H_0\ll M_{\rm P}$ (because gravity must be in its perturbative regime).

 Therefore, we need to look harder for an interaction that works. One possible way out is to admit non-canonical couplings. 
Taking, for example,
$G(\phi)=(M_{\rm P}/\phi)^{2n}$  in~(\ref{typical interactions: Einstein frame:2}) results in the photon effective cubic coupling,
$\sigma_{\rm eff} \sim k^2/M_{\rm P}(H_0/H_E)^n$. This process is kinematically allowed when the photon momenta 
$k<\omega_E\sim H_E$, and therefore,
$\sigma_{\rm eff} \sim H_0^n/(M_{\rm P} H_E^{n-2})$. The corresponding decay rate is then, 
$\Gamma\sim H_0^{2n}/(M_{\rm P}^2H_E^{2n-3})$. The rate must grow with respect to $H_E$, such that eventually,
 $\Gamma/H_E\sim H_0^{2n}/(M_{\rm P}^2H_E^{2(n-1)})\gg 1$, implying that $n>1$; the simplest case when this is realised is when 
$n=3/2$ and $n=2$, for which we require $H_E/H_0\ll (H_0/M_{\rm P})^2$ and  $H_E/H_0\ll (H_0/M_{\rm P})^2$, respectively, 
which can be satisfied for a sufficiently small $H_E$. One can show that attempting to introduce a power-law 
nonminimal coupling into the Yukawa and $\phi-\chi$ interaction does not open any kinematically allowed decay channel.

 But can we construct a decay with canonical couplings that works? Let us consider the following simple bilinear Lagrangian density, 
\begin{equation}
   \sqrt{-g}{\cal L}^\prime _{\rm int} = -\sqrt{-g}\frac12 m_{\phi\chi}^2\phi\chi \;\Longrightarrow \;
  {\cal L}^\prime _{E\rm int} 
 = -\frac{M_{\rm P}}2 \frac{m_{\phi\chi}^2}{F^{3/2}}\phi(\phi_E)\chi_E 
  \rightarrow \frac12 m_{\phi\chi}^2 {\exp}\left(-\frac{\lambda_E\phi_E}{2M_{\rm P}}\right)\chi_E
\,.
\label{interaction lagrangian: simple}
\end{equation}
When the cubic interaction is extracted, one gets a decay, $\phi\rightarrow \phi\chi$, 
but this channel has a very small (classically zero) phase space, and hence we shall neglect it. On the other 
hand, $\phi$ will constantly convert into $\chi$ through the bi-linear coupling term (mass oscillations).
In flat space these (neutrino-like) mass oscillations do not induce any decays, because they are reversible 
(formally, the decay rate for that process is zero). However, in an expanding universe, the created $\chi$ particles
are massless, and their energy density redshifts as $\rho_\chi\propto 1/a^4$ (the corresponding 
$\epsilon=2$), such that the process is not any more reversible: more energy gets transferred 
from $\phi$ into $\chi$ than {\it v.v}, and one effectively gets a decay. The decay is for sure 
kinematically allowed (because $\chi$ is massless); 
the question is whether this decay is sufficiently effective. The decay rate can be estimated 
as follows, $\Gamma\sim m_{\phi\chi}^4(H_E/H_0)^2/[\|\vec k\|^2 H_E]\gg H_E$.
With $\|\vec k\,\|_{\rm max}\sim H_E$ one gets that the process becomes fast when $H_E/H_0\ll (m_{\phi\chi}/H_0)^2$.
Note that, when one includes gravity, 
this process can be viewed as a perturbative decay of $\phi$ into one $\chi$ and one graviton.

\end{document}